\documentclass[twocolumn]{article}

\usepackage[english]{babel}
\usepackage[numbers]{natbib}

\usepackage{longtable} 
\usepackage{array} 
\usepackage{lscape} 
\usepackage{multirow}

\usepackage[letterpaper,top=2cm,bottom=2cm,left=3cm,right=3cm,marginparwidth=1.75cm]{geometry}

\usepackage{amsmath}
\usepackage{graphicx}
\PassOptionsToPackage{colorlinks=true, allcolors=blue}{hyperref}
\usepackage{hyperref}

\date{}
\title{Mapping Executive Function Tasks for Children: A Scoping Review for Designing a Research-Oriented Platform}

\author{
    Matheus Rodrigues Felizardo\textsuperscript{1} \\ \small\texttt{up202300391@up.pt} \\
    António Coelho\textsuperscript{1} \\ \small\texttt{acoelho@fe.up.pt} 
    \and
    Nuno Miguel Feixa Rodrigues\textsuperscript{2} \\ \small\texttt{nfr@ipca.pt} \\
    Eva Ferreira de Oliveira\textsuperscript{2} \\ \small\texttt{eoliveira@ipca.pt} 
    \and
    Sónia Silva Sousa\textsuperscript{3} \\ \small\texttt{soniamachado@psi.uminho.pt} \\
    Adriana Sampaio\textsuperscript{3} \\ \small\texttt{adriana.sampaio@psi.uminho.pt}
}

\date{
\textsuperscript{1}Porto University\\
\textsuperscript{2}Polytechnic University of Cávado and Ave\\
\textsuperscript{3}University of Minho
}

\begin{document}
\maketitle

\begin{abstract}
\textbf{BACKGROUND:} Executive functions (EFs) are a set of cognitive processes with inhibition, working memory, and cognitive flexibility as their core components. These processes help individuals control impulses, stay focused, think before acting, and mentally handle information. Childhood is a critical period for the development of EF, and effective assessment and intervention tools play a key role in the development of these skills. However, there are few standardized tools, and not many studies combine EF tasks with physical activity in a gamified approach.

\textbf{OBJECTIVES:} This scoping review aims to map the existing EF tasks used for children, identify common implementation strategies, and explore methods for measuring outcomes, providing a foundation for the development of a research-oriented and structured platform to assess executive functions in childhood.

\textbf{DESIGN:} A systematic search was conducted in SCOPUS, ScienceDirect, and ERIC databases using the query "executive function task" AND (children OR child OR childhood)". Inclusion criteria included studies published between 2019 and 2024, written in English, with participants aged 5 to 9 years. Extracted data included details on stimuli, trials, measures, scoring mechanisms, and stop conditions. Studies were excluded if they lacked clear methodological descriptions, such as undefined stopping criteria or unspecified task adaptation methods, which compromised their comparability.

\textbf{RESULTS:} A total of 2044 articles were identified, with 113 duplicates removed. After selection and eligibility evaluation, 23 studies met the inclusion criteria. The list of identified tasks can be found in the table~\ref{tab:summary_ef_tasks}. The findings highlight key EF tasks, implementation strategies, and measurement methodologies.

\textbf{CONCLUSIONS:} Integrating EF tasks into a structured platform for children offers a promising approach to address research gaps, standardize assessments, and provide researchers with a flexible and reliable tool for studying EF development.

\end{abstract}

\section{Introduction}
Executive functions (EFs) are a set of cognitive processes responsible for \textbf{inhibition}, \textbf{working memory}, and \textbf{cognitive flexibility}. These abilities are crucial for children’s \textit{academic success}, \textit{social interactions}, and \textit{emotional regulation}~\cite{diamond2013executive}. Childhood represents a \textit{critical period for EF development}, with significant growth occurring between the ages of \textbf{5 and 9 years}~\cite{miyake2012individual}. Assessing and strengthening executive functions in school-aged children is essential for promoting academic achievement and reducing educational disparities. Research has consistently demonstrated that EF skills correlate with mathematics, reading comprehension, and writing proficiency, sometimes even more strongly than IQ~\cite{shaul2013therole}.

Recent studies highlight the importance of using fun and age-appropriate tasks to assess and improve EF skills in children~\cite{best2010developmental}. However, several challenges persist. Traditional methods, such as paper-based neuropsychological tests and teacher/parent rating scales, often lack real-world applicability and cultural adaptability, making them less effective in capturing children's everyday EF-related challenges~\cite{fahy2014assessment}. 

In digital assessments, some studies \textbf{lack clarity in defining trial parameters}, such as the number of repetitions or stopping conditions, limiting their applicability across diverse contexts. Despite strong evidence supporting the benefits of physical activity for cognitive development, integration into EF tasks remains limited~\cite{tomporowski2015exercise}. Additionally, there is a lack of standardized, scalable, and easily accessible tools for monitoring EF development over time.

To address these challenges, there is a growing interest in digital platforms that allow real-time, research-oriented EF assessment. Digital solutions provide advantages such as dynamic task adaptation, automated data collection, and greater accessibility across diverse educational settings, particularly for children with low levels of text comprehension~\cite{ruffini2023whichef}. Digital EF assessments enable the creation of new, game-based task implementations with customizable parameters. By allowing dynamic adjustments to stimuli, content, and difficulty levels, these systems enhance accuracy, engagement, and ecological validity. Such adaptability ensures that EF tasks can be tailored to different research needs, educational contexts, and individual cognitive profiles, making assessments more flexible, scalable, and accessible.

This scoping review aims to \textbf{map the existing EF tasks used for children}, identify \textbf{common implementation strategies}, and explore \textbf{methods for measuring outcomes}, providing a foundation for designing a structured, adaptable, and evidence-based tool that facilitates standardized EF assessments and supports educational and cognitive interventions.

\section{Methods}
This scoping review follows the PRISMA-ScR guidelines for scoping reviews.

\subsection{Target Population}
The target population consists of \textbf{children aged 5 to 9 years}. This range represents a \textit{critical period} for the acquisition and refinement of \textbf{inhibition}, \textbf{working memory}, and \textbf{cognitive flexibility}, aligning with key \textit{educational milestones} from \textit{preschool to early primary education}~\cite{diamond2013executive, best2010developmental}.

\subsection{Research Questions}
\begin{enumerate}
    \item \textbf{What types of executive function tasks are commonly used for children?}
    \item \textbf{What are the main strategies for applying these tasks to children?}
    \item \textbf{How are these tasks used to measure and enhance executive functions in children?}
\end{enumerate}

\subsection{Search Strategy and Study Selection}

The search strategy was designed to identify studies published between \textbf{2019 and 2024}. This time frame was chosen to ensure that the review captures the most recent developments in executive function (EF) assessment methods while maintaining a manageable scope, allowing for a more feasible data collection process, and still ensuring the inclusion of studies reflecting current research trends. Additionally, as digital and gamified EF assessments have evolved rapidly in recent years, focusing on the most recent literature helps highlight the latest advancements in the field.

\begin{figure}[h!]
    \centering
    \includegraphics[width=1\linewidth]{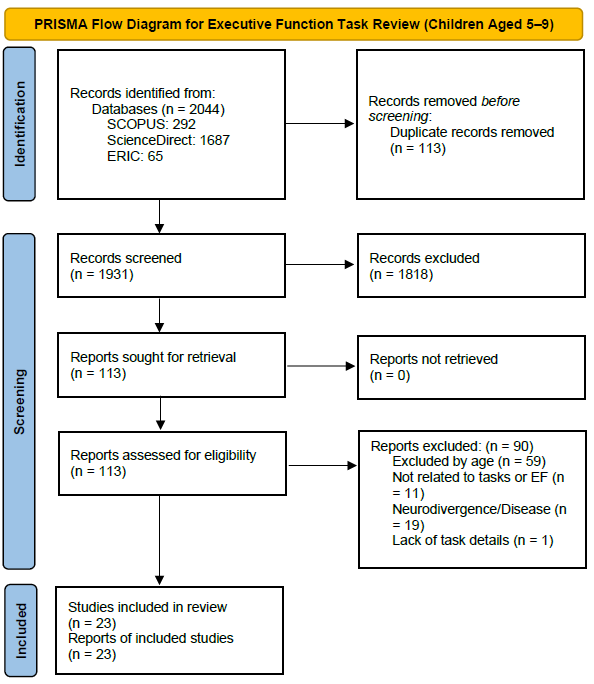}
    \caption{PRISMA Flow Diagram for Executive Function Task Review (Children Aged 5–9).}
    \label{fig:flow-diagram}
\end{figure}

The study selection process adhered to PRISMA guidelines. The flow chart (Figure~\ref{fig:flow-diagram}) details the stages from identification to inclusion.

\subsection{Data Extraction}
Data extraction focused on:
\begin{itemize}
    \item \textbf{Study Characteristics:} Author(s), year, and design.
    \item \textbf{Population Details:} Age range.
    \item \textbf{Task Details:} Task type, target EF, objectives.
    \item \textbf{Task Implementation:} Stimuli, trials, measures, scoring, stop conditions.
\end{itemize}

\section{Results}
\subsection{Executive Function Tasks Identified}

The selected studies included multiple executive function tasks. Many of the articles provided rich details about the tasks implementation and many of them used more than one tasks in the study.

Below, in the table~\ref{tab:summary_ef_tasks}, each study is listed along with the tasks name, their objectives, stimuli, and some implementation details.  

The studies reviewed reveal a variety of tasks used to assess core executive functions in children. These tasks measure aspects such as inhibition, working memory, and cognitive flexibility. For instance, tasks like the Flanker and Go/No-Go Task primarily assess inhibition, requiring participants to focus on specific targets while ignoring distractors or suppressing responses to certain stimuli. These tasks are typically implemented using visual stimuli, such as arrows or animals, presented on a computer screen or physical cards, and include a series of trials with varying difficulty levels.

For working memory, tasks like the Backward Digit Span and the Corsi Blocks Backward Task were common, requiring children to recall sequences in reverse order. These tasks use auditory or visual sequences and challenge children by increasing sequence length as they progress. The Dimensional Change Card Sort (DCCS) task is key for assessing cognitive flexibility. It requires children to switch sorting rules based on color or shape, testing their ability to adapt to changing rules.

Many tasks use block-based designs, dividing trials into practice, test, and mixed sections, with increasing difficulty. These designs assess baseline performance while challenging children with variations in stimuli to test adaptability and learning abilities.

Most studies employed visual stimuli, with some using auditory cues or multisensory elements to engage children. The variety of stimuli types demonstrates the flexibility of these tasks in assessing different executive functions and how sensory modalities may influence performance.

Even with different methods, common measures were used, including reaction time (RT), accuracy, and performance metrics like error rates or interference scores. These measures provide valuable insights into the child’s cognitive abilities and their management of task demands and distractions.

\subsection{Summary of Key Findings}
This scoping review provides a comprehensive analysis of the executive function (EF) tasks used in children aged 5–9 years, highlighting the diversity of tasks across studies. A total of 28 distinct EF tasks were identified, with several variations in their implementation.

Cognitive flexibility tasks, such as the Dimensional Change Card Sort (DCCS) and the Wisconsin Card Sorting Task (WCST), assess a child’s ability to shift between rules and adapt thinking. Inhibitory control was primarily assessed using the Flanker Task and Go/No-Go Task, both measuring the ability to focus attention and suppress responses to irrelevant stimuli. These tasks often had slight variations in stimuli, such as arrows or fish, but shared the core goal of measuring inhibition.

Working memory tasks, like the Backward Digit Span and Working Memory Task, typically involved recalling sequences of numbers or visual locations, assessing short-term memory and the ability to manipulate information. Additionally, tasks like the Head-Toes-Knees-Shoulders (HTKS) and DCCS measured both inhibition and cognitive flexibility.

Many studies employed tasks with mixed cognitive demands, such as those that combined cognitive flexibility, inhibition, and working memory, highlighting the interconnected nature of these functions. Tasks like the Stroop Task had different versions, but all aimed to measure a child’s ability to control automatic responses.

Stimuli were diverse, with visual cues such as arrows and animals being common, though auditory signals were also used. The variety of stimuli shows the adaptability of EF tasks to engage children with different learning styles and ensure reliable results.

In summary, the review identifies the most common EF tasks for children aged 5–9 years, emphasizing that small variations in task design can assess similar cognitive processes like inhibition, cognitive flexibility, and working memory.

\section{Limitations}
This review has some limitations that should be acknowledged. 

One of the main constraints concerns the age range criterion (5–9 years) used for study inclusion. This restriction may have excluded studies with executive function tasks aimed at younger children, particularly those designed for 4-year-olds. Such tasks might have provided valuable insights into developmental transitions or specific adaptations necessary for creating age-appropriate tools and platforms.

Additionally, the review focused on studies published between 2019 and 2024. While this approach ensured an up-to-date analysis, it may have unintentionally excluded older yet still relevant research that could offer important theoretical or methodological contributions. However, these decisions were essential to maintain a manageable scope and enable a thorough analysis within the available resources.

\section{Discussion}

The findings from this scoping review highlight both the progress and persistent challenges in the assessment of executive functions (EF) in children. While a variety of tasks have been developed to measure inhibition, working memory, and cognitive flexibility, substantial inconsistencies exist in their \textbf{implementation, trial parameters, and scoring methods}. These discrepancies pose significant barriers to the \textbf{comparability of results across studies}, ultimately limiting the field’s ability to establish robust conclusions regarding EF development.

To better illustrate the landscape of EF assessments, Table~\ref{tab:summary_ef_tasks_classified} provides a structured classification of tasks according to their primary EF component—Inhibitory Control, Working Memory, or Cognitive Flexibility. This categorization highlights the diversity of tasks used in research and reveals an uneven distribution of EF components, with some functions receiving significantly more focus than others. Furthermore, Table~\ref{tab:summary_ef_tasks} presents a comprehensive synthesis of EF tasks, detailing their methodologies, implementation strategies, and scoring mechanisms, reinforcing the methodological inconsistencies across studies. These insights emphasize the need for a more standardized and structured approach to EF task design and implementation.

\subsection{Lack of Standardization Across Studies}
One of the main issues identified is the \textbf{lack of consistency in task parameters}, even for widely used assessments. Tasks such as the \textit{Flanker Task}, \textit{Go/No-Go Task}, and \textit{Dimensional Change Card Sort (DCCS)} exhibit considerable \textbf{variations in trial lengths, block structures, and response criteria}, which directly impact the difficulty and interpretation of results. For example, in the \textit{Go/No-Go Task}, the proportion of "go" versus "no-go" trials varies across studies, affecting inhibitory demands and making comparisons difficult. Similarly, in the \textit{Flanker Task}, discrepancies in stimulus presentation times and response deadlines lead to inconsistent difficulty levels across implementations.

Beyond these technical aspects, scoring criteria also differ widely. Some tasks, particularly those embedded in standardized cognitive batteries (e.g., \textit{Backward Digit Span}), benefit from well-established norms, allowing results to be contextualized within broader developmental frameworks. However, many other EF tasks lack reference data, making it unclear \textbf{how performance should be interpreted across different samples and cultural settings}. Without standardized guidelines, results remain difficult to compare, limiting the generalizability of EF research.

\subsection{The Need for a Unified Framework}
Given these challenges, a structured and \textbf{standardized digital platform} is necessary to enhance the comparability and reliability of EF assessments. Such a platform would serve to:
\begin{itemize}
    \item \textbf{Standardize task implementation} – Define fixed parameters for trial numbers, timing, and scoring to ensure consistency across studies.
    \item \textbf{Ensure cross-cultural applicability} – Adapt tasks to different linguistic and educational contexts while maintaining their core cognitive demands.
    \item \textbf{Facilitate large-scale data collection} – Allow researchers to gather and compare EF performance across diverse populations with greater methodological rigor.
    \item \textbf{Integrate physical activity into digital assessments} – Many EF tasks remain \textbf{static}, despite evidence that \textbf{motor engagement enhances cognitive performance}~\cite{tomporowski2015exercise}. A modern framework should explore \textbf{ways to incorporate movement-based tasks digitally}, using motion tracking, adaptive feedback, or hybrid digital-physical designs.
\end{itemize}

Tasks that are highly structured and \textbf{less dependent on environmental context}, such as the \textit{Flanker Task}, \textit{DCCS}, and \textit{Go/No-Go Task}, are well-suited for digital adaptation and could serve as the foundation for a standardized framework. In contrast, \textbf{movement-based EF tasks}, such as \textit{HTKS} and \textit{Freeze Task}, require innovative \textbf{digital adaptations} to preserve their cognitive and motor demands while ensuring consistency across studies.

Additionally, some tasks present specific characteristics that make them particularly promising for digital adaptation. For \textbf{inhibitory control}, tasks such as the \textit{Stroop Task}, \textit{Flanker Task}, and \textit{Go/No-Go Task} have well-established methodologies and are widely implemented in computerized settings. In the case of \textbf{working memory}, tasks like the \textit{Backward Digit Span}, \textit{Corsi Blocks Task}, and \textit{Keep Track Task} provide reliable cognitive measures and can be effectively implemented in digital formats with minimal modifications. Finally, for \textbf{cognitive flexibility}, the \textit{Dimensional Change Card Sort (DCCS)}, \textit{Wisconsin Card Sorting Task (WCST)}, and \textit{Dots Task} offer strong candidates for digital transformation due to their structured nature and adaptability to automated scoring.

\subsection{Conclusion}
The extensive \textbf{variability in EF task design and administration} underscores the urgent need for a \textbf{standardized framework} to improve research reliability and applicability. Furthermore, the \textbf{underutilization of movement-based tasks}, despite their \textbf{proven cognitive benefits}, highlights an area for innovation in EF assessment. The development of a \textbf{research-oriented digital platform} offers a \textbf{practical solution} to these methodological inconsistencies, ensuring that EF assessments are comparable, replicable, and accessible across diverse research contexts. By integrating \textbf{well-defined parameters, standardized scoring mechanisms, and adaptable task designs—including movement-based elements in a digital format}—this platform has the potential to \textbf{bridge existing gaps} in EF assessment and advance our understanding of cognitive development in children.

\bibliographystyle{unsrtnat}

\appendix
\section*{Appendix}

\clearpage
\onecolumn
\begin{longtable}{|p{3cm}|p{3cm}|p{1cm}|p{3cm}|p{5cm}|}
\caption{Executive Function Tasks Categorized by EF Component}
\label{tab:summary_ef_tasks_classified} \\

\hline
\textbf{Task} & \textbf{Author(s)} & \textbf{Year} & \textbf{Age Range/Mean} & \textbf{Notes} \\ 
\hline
\endfirsthead

\hline
\textbf{Task} & \textbf{Author(s)} & \textbf{Year} & \textbf{Age Range/Mean} & \textbf{Notes} \\ 
\hline
\endhead

\hline
\multicolumn{5}{|r|}{\textit{Continued on next page}} \\ 
\hline
\endfoot

\hline
\endlastfoot

\multicolumn{5}{|c|}{\textbf{Inhibitory Control - 18 tasks}} \\ 
\hline

\multirow{7}{*}{Stroop Task \textbf{(7)}} 
    & Bellon et al.~\cite{bellon_more_2019} & 2019 & 6-8 years & Animal Stroop Task  \\ 
    & Lertladaluck et al.~\cite{lertladaluck_executive_2024} & 2024 & 5–7 years &  \\ 
    & Lin et al.~\cite{lin_measurement_2019} & 2019 & \ensuremath{\approx} 5 years & Shape Stroop Task  \\ 
    & Papastergiou et al.~\cite{papastergiou_study_2022} & 2022 & 5–9 years & Nonverbal Stroop Task \\ 
    & Schirmbeck et al.~\cite{schirmbeck_contrasting_2021} & 2021 &  8–10 years & Child-Friendly Stroop Task \\ 
    & Veneziano et al.~\cite{veneziano_individual_2022} & 2022 & 6–-8 years & Animal Stroop Task\\ 
    & Zanto et al.~\cite{zanto_digital_2024} & 2024 & 8–9 years & \\ 
\hline

\multirow{6}{*}{Flanker Task \textbf{(6)}} 
    & Bellon et al.~\cite{bellon_more_2019}{*} & 2019 & 6–-8 years &\\ 
    & Grenell et al.~\cite{grenell_childrens_2024} & 2024 & 7–9 years &\\ 
    & Lertladaluck et al.~\cite{lertladaluck_executive_2024} & 2024 & 5–7 years &\\ 
    & Maurer et al.~\cite{maurer_towards_2019} & 2019 & 5–6 years &\\ 
    & Oeri et al.~\cite{oeri_regulating_2020} & 2020 & 5–6 years & \\ 
    & Zanto et al.~\cite{zanto_digital_2024} & 2024 & 8–9 years &\\ 
\hline

\multirow{3}{*}{Simon Says Task \textbf{(3)}} 
    & Keşşafoğlu et al.~\cite{kessafoglu_immediate_2024} & 2024 & 5–6 years & \\
    & Lavis et al.~\cite{lavis_ill_2021} & 2021 & \ensuremath{\approx} 5 years &  \\ 
    & Traverso et al.~\cite{traverso_relationship_2021} & 2021 & 5–6 years &  \\ 
\hline

\multirow{3}{*}{Go/No-Go Task \textbf{(3)}} 
    & Schäfer et al.~\cite{schafer_executive_2024} & 2024 & 6–-8 years & \\ 
    & Xie et al.~\cite{xie_using_2024} & 2024 & 5–6 years & \\ 
    & Xie et al.~\cite{xie_enhancing_2022} & 2022 & 5–6 years & \\ 
\hline

\multirow{2}{3cm}{Hearts and Flowers Task \textbf{(2)}} 
    & Ger et al.~\cite{ger_is_2024} & 2024 & 6–-8 years & This task also involves cognitive flexibility.\\ 
    & Ger et al.~\cite{ger_monitoring_2024} & 2024 & 6–-8 years &  \\ 
\hline

\multirow{2}{3cm}{Head-Toes-Knees-Shoulders \textbf{(2)}} 
    & Ahmed et al.~\cite{ahmed_cognition_2021} & 2020 & 5–6 years & This task also involves working memory.\\ 
    & Kwakkel et al.~\cite{kwakkel_impact_2021} & 2021 & 5–6 years &  \\ 
\hline

\multirow{1}{*}{Freeze Task \textbf{(1)}} 
    & Ahmed et al.~\cite{ahmed_cognition_2021} & 2020 & 5–6 years & \\ 
\hline

\multirow{1}{*}{Jumping Task \textbf{(1)}} 
    & Ahmed et al.~\cite{ahmed_cognition_2021} & 2020 & 5–6 years & \\ 
\hline

Pair Cancellation Task \textbf{(1)} 
    & Ahmed et al.~\cite{ahmed_cognition_2021} & 2020 & 5--6 years & \\ 
\hline

Red/Blue Task \textbf{(1)}
    & Ishikawa et al.~\cite{ishikawa_relationship_2023} & 2023 & 5–6 years &  \\
\hline

Bear/Dragon \textbf{(1)}
    & Ishikawa et al.~\cite{ishikawa_relationship_2023} & 2023 & 5–6 years &  \\
\hline

Continuous Performance Task \textbf{(1)}
    & Kwakkel et al.~\cite{kwakkel_impact_2021} & 2021 & 5–6 years &\\ 
\hline

Luria Hand Game \textbf{(1)}
    & Lertladaluck et al.~\cite{lertladaluck_executive_2024} & 2024 & 5–7 years & \\ 
\hline

Snack Delay Task \textbf{(1)} 
    & Lin et al.~\cite{lin_measurement_2019} & 2019 & \ensuremath{\approx} 5 years & \\ 
\hline

Toy Delay Task \textbf{(1)} 
    & Lin et al.~\cite{lin_measurement_2019} & 2019 & \ensuremath{\approx} 5 years & \\ 
\hline

Preschool Matching Familiar Figure Task \textbf{(1)} 
    & Traverso et al.~\cite{traverso_relationship_2021} & 2021 & 5–6 years &  \\ 
\hline

Stop Signal Task \textbf{(1)} 
    & Zhang et al.~\cite{zhang_recreational_nodate} & 2024 & 5–6 years & \\ 
\hline

Marching Task \textbf{(1)} 
    & Ahmed et al.~\cite{ahmed_cognition_2021} & 2020 & 5–6 years & \\ 
\hline

\multicolumn{5}{|c|}{\textbf{Working Memory - 12 tasks}} \\ 
\hline

\multirow{4}{3cm}{Backward Digit/Word Span \textbf{(4)}} 
    & Ahmed et al.~\cite{ahmed_cognition_2021} & 2020 & 5–6 years & \\ 
    & Papastergiou et al.~\cite{papastergiou_study_2022} & 2022 & 5–9 years & \\ 
    & Keşşafoğlu et al.~\cite{kessafoglu_immediate_2024} & 2024 & 5–6 years & \\ 
    & Lavis et al.~\cite{lavis_ill_2021} & 2021 & \ensuremath{\approx} 5 years & \\
\hline

\multirow{2}{3cm}{Working Memory Task \textbf{(2)}} 
    & Xie et al.~\cite{xie_using_2024} & 2024 & 5–6 years & \\ 
    & Zanto et al.~\cite{zanto_digital_2024} & 2024 & 8–9 years & \\ 
\hline

\multirow{2}{3cm}{Keep Track Task \textbf{(2)}} 
    & Traverso et al.~\cite{traverso_relationship_2021} & 2021 & 5–6 years & \\ 
    & Zhang et al.~\cite{zhang_recreational_nodate} & 2024 & 5–6 years & \\ 
\hline

Forward Word Span \textbf{(1)}
    & Kwakkel et al.~\cite{kwakkel_impact_2021} & 2021 & 5–6 years & \\ 
\hline

Animal Updating Task \textbf{(1)} 
    & Maurer et al.~\cite{maurer_towards_2019} & 2019 & 5–6 years & \\ 
\hline

2-Back Task \textbf{(1)} 
    & Bellon et al.~\cite{bellon_more_2019} & 2019 & 6–-8 years & \\ 
\hline

Object Span Task \textbf{(1)} 
    & Schirmbeck et al.~\cite{schirmbeck_contrasting_2021} & 2021 & 8–10 years & \\ 
\hline

Corsi Blocks Backward Task \textbf{(1)} 
    & Schäfer et al.~\cite{schafer_executive_2024} & 2024 & 6–-8 years & \\ 
\hline

Dual Request Selective Task \textbf{(1)} 
    & Traverso et al.~\cite{traverso_relationship_2021} & 2021 & 5–6 years & \\ 
\hline

Missing Scan Task \textbf{(1)} 
    & Xie et al.~\cite{xie_enhancing_2022} & 2022 & 5–6 years & \\ 
\hline

Letter Memory Task \textbf{(1)} 
    & Zhang et al.~\cite{zhang_recreational_nodate} & 2024 & 5–6 years & \\ 
\hline

List Sorting Task \textbf{(1)} 
    & Grenell et al.~\cite{grenell_childrens_2024} & 2024 & 7–9 years &\\  
\hline

\multicolumn{5}{|c|}{\textbf{Cognitive Flexibility - 8 tasks}} \\ 
\hline

\multirow{5}{3cm}{Dimensional Change Card Sort (DCCS) \textbf{(5)}} 
    & Grenell et al.~\cite{grenell_childrens_2024} & 2024 & 7–9 years & \\ 
    & Ishikawa et al.~\cite{ishikawa_relationship_2023} & 2023 & 5–6 years & \\ 
    & Maurer et al.~\cite{maurer_towards_2019} & 2019 & 5–6 years &  \\ 
    & Xie et al.~\cite{xie_using_2024} & 2024 & 5–6 years & \\ 
    & Xie et al.~\cite{xie_enhancing_2022} & 2022 & 5–6 years & \\ 
\hline

\multirow{2}{3cm}{Wisconsin Card Sorting Task (WCST) \textbf{(2)}} 
    & Bellon et al.~\cite{bellon_more_2019} & 2019 & 6–-8 years & \\ 
    & Chan et al.~\cite{chan_structure-function_2022} & 2022 & \ensuremath{\approx} 5 years & \\ 
\hline

\multirow{2}{3cm}{Flexible Item Selection \textbf{(2)}} 
    & Keşşafoğlu et al.~\cite{kessafoglu_immediate_2024} & 2024 & 5–6 years & \\ 
    & Schäfer et al.~\cite{schafer_executive_2024} & 2024 & 6–-8 years & \\ 
\hline

Colour-Shape Task \textbf{(1)}
    & Papastergiou et al.~\cite{papastergiou_study_2022} & 2022 & 5–9 years & \\ 
\hline

Classification Task \textbf{(1)} 
    & Veneziano et al.~\cite{veneziano_individual_2022} & 2022 & 6–-8 years & \\ 
\hline

Local/Global Task \textbf{(1)} 
    & Veneziano et al.~\cite{veneziano_individual_2022} & 2022 & 6–-8 years & \\ 
\hline

Dots Task \textbf{(1)} 
    & Zhang et al.~\cite{zhang_recreational_nodate} & 2024 & 5–6 years & \\ 
\hline

Minnesota Executive Function Scale (MEFS) \textbf{(1)} 
    & Oeri et al.~\cite{oeri_regulating_2020} & 2020 & 5–6 years & This task also involves working memory and inhibitory control. \\ 
\hline
\end{longtable}

\begin{landscape}
\begin{longtable}{|p{2.5cm}|p{1.5cm}|p{2.5cm}|p{2.5cm}|p{4cm}|p{4cm}|p{3.5cm}|}
\caption{Summary of Executive Function Tasks from Included Studies (5–9 years)}
\label{tab:summary_ef_tasks} \\
\hline
\textbf{Author(s)} & \textbf{Year} & \textbf{Task} & \textbf{Target EF} & \textbf{Objective} & \textbf{Stimulus} & \textbf{Trials} \\ 
\hline
\endfirsthead

\hline
\textbf{Author(s)} & \textbf{Year} & \textbf{Task} & \textbf{Target EF} & \textbf{Objective} & \textbf{Stimulus} & \textbf{Trials} \\ 
\hline
\endhead

\hline
\multicolumn{7}{|r|}{\textit{Continued on next page}} \\ 
\hline
\endfoot

\hline
\endlastfoot

Ahmed et al.~\cite{ahmed_cognition_2021} & 2020 & Freeze Task & Inhibitory Control & Children march in a circle and freeze when music starts. They unfreeze when instructed or when music stops. & Music played for random intervals (7–15 seconds) & 5 trials (Stop time: Reaction to freezing and distractibility) \\ 
 & & Jumping Task & Inhibitory Control, Working Memory & Children follow one- to three-step instructions after freezing (e.g., jump three times, clap twice). & Music played for random intervals (7–15 seconds) & 3 trials (Stop time, distractibility, action recall, action performance) \\ 
 & & Marching Task & Inhibitory Control, Cognitive Flexibility & Children march in one direction for a specific song and the opposite direction for another song. & Music played for random intervals (7–15 seconds) & 7 trials (Stop time, distractibility, march recall, march performance) \\ 
 & & Head-Toes-Knees-Shoulders & Inhibitory Control & Children perform the opposite of the instructed movement (e.g., touch toes when asked to touch head). & Not specified & 30 trials across 3 blocks \\ 
 & & Backward Digit Span & Working Memory & Children repeat numbers in reverse order. & Spoken sequences of numbers & Not specified \\ 
 & & Pair Cancellation Task & Inhibitory Control & Children identify specific pairs (e.g., a ball followed by a dog) on a stimulus sheet. & Rows of pictures on a sheet & Not specified \\ 
\hline

Bellon et al.~\cite{bellon_more_2019} & 2019 & Flanker Task & Inhibitory Control & Assess inhibition by responding to a central arrow’s direction while ignoring flanking distractors. & Arrows flanked by distractors, presented for 2500 ms and black screen visible for 1000 ms & 40 test trials (50\% incongruent), preceded by 12 baseline trials with single arrows. \\ 
 &  & Animal Stroop Task & Inhibitory Control & Assess inhibition by identifying the larger animal in real life, ignoring image size. & Two animals presented simultaneously, with one image four times larger. Stimuli appeared for 2000 ms, followed by a black screen that remained visible for 1000 ms & 40 test trials (50\% incongruent), preceded by 20 baseline trials with equal-sized images. \\ 
 &  & Wisconsin Card Sorting Task (WCST) & Cognitive Flexibility & Measure cognitive flexibility by sorting cards based on changing rules (color or shape). & Cards with specific colors and shapes. & 50 trials; sorting rule changes after 7–9 correct responses. \\ 
 &  & 2-Back Task & Working Memory & Measure updating by identifying if a current stimulus matches one shown two trials earlier. & Colored images, presented for 3000 ms each followed by black screen that remained visible for 1000 ms & 40 test trials, with 30\% being targets. \\ \hline

Chan et al.~\cite{chan_structure-function_2022} & 2022 & Wisconsin Card Sorting Test (WCST) & Cognitive Flexibility & Measure cognitive flexibility by switching rules in categorization tasks. & Cards with attributes of color, shape, and number; rules change after 10 consecutive correct responses. & 64 total trials. \\ \hline

Ger et al.~\cite{ger_is_2024} & 2024 & Hearts and Flowers Task & Inhibitory Control, Cognitive flexibility & Assess inhibition and cognitive flexibility through rule-switching. & Geometric shapes (hearts and flowers); heart = congruent, flower = incongruent. & Hearts: 24, Flowers: 36, Mixed: 48 (with 12 flowers). \\ \hline

Ger et al.~\cite{ger_monitoring_2024} & 2024 & Hearts and Flowers Task & Inhibitory Control, Cognitive Flexibility & Assess inhibitory control and cognitive flexibility. & Stimuli presented on a screen: Hearts (congruent) and Flowers (incongruent). Each trial has 750 ms stimulus duration, followed by a 500 ms fixation cross. & Block 1: 24 congruent trials; Block 2: 36 incongruent trials; Block 3: 48 congruent + 12 incongruent trials. \\ \hline

Grenell et al.~\cite{grenell_childrens_2024} & 2024 & Flanker Task & Inhibitory Control & Measure inhibitory control and attention. & Five fish or arrows; central target’s direction is identified (e.g., <<<<< or <<><<). & Not specified; includes congruent and incongruent trials. \\ 
 & & List Sorting Task & Working Memory & Measure working memory by recalling sequences of items by category and size. & Visual presentation of items (e.g., animals, foods). & 6 levels (2–7 items). \\ 
 & & Dimensional Change Card Sort (DCCS) & Cognitive Flexibility & Measure cognitive flexibility by switching between sorting by shape or color. & Cards with objects of varying shapes and colors. & Practice trials followed by test trials (number not specified). \\ \hline

Ishikawa et al.~\cite{ishikawa_relationship_2023} & 2023 & Red/Blue Task & Inhibitory Control & Measure inhibitory control through responses to incongruent color names. & Cards colored red or blue; children must point to the opposite color when named by the experimenter. & 10 test trials (5 red, 5 blue). \\ 
 & & Bear/Dragon Task & Inhibitory Control & Measure inhibitory control by following commands from the "bear" and inhibiting those from the "dragon". & Commands issued by two puppets (bear and dragon). & 12 trials (6 per puppet). \\ 
 & & DCCS Task & Cognitive Flexibility & Assess cognitive flexibility through rule-switching in card sorting. & Cards with two dimensions (color and shape); sorted based on alternating rules. & 8 trials per phase (Pre-, Post-Switch, Mixed). \\ \hline

 Keşşafoğlu et al.~\cite{kessafoglu_immediate_2024} & 2024 & Backward Word Span & Working Memory & Measure working memory by recalling words in reverse order. & Verbal sequences of 2–5 words (e.g., "curtain, garden"). & 8 test trials. \\ 
 & & Simon Says Task & Inhibitory Control & Assess inhibitory control via selective imitation and non-imitation of commands. & Actions (e.g., touch your nose) given with or without "Simon Says." & 10 trials (5 imitation, 5 anti-imitation). \\ 
 & & Flexible Item Selection & Cognitive Flexibility & Measure cognitive flexibility by selecting cards matching on varying dimensions. & Three cards per trial; select two matching on one dimension, then another. & 15 trials. \\ \hline

Kwakkel et al.~\cite{kwakkel_impact_2021} & 2021 & Head-Toes-Knees-Shoulders (HTKS) & Inhibitory Control, Working Memory & Measure self-regulation by responding oppositely to commands. & Verbal commands like "Touch your head"; responses must be opposite (e.g., touch toes). & 20 test items (10 per rule set). \\ 
 & & Continuous Performance Task (CPT) & Inhibitory Control & Measure sustained attention using go/no-go responses. & Visual task: target = red circle, non-target = red square; auditory task: high/low tones. & 200 trials per task (80 target, 120 non-target). \\ 
 & & Forward Word Span & Working Memory & Measure working memory by recalling word sequences. & Spoken words in sequences of 2–7. & 2 trials per sequence length. \\ \hline

 Lavis et al.~\cite{lavis_ill_2021} & 2021 & Simon Says Task & Inhibitory Control & Assess inhibition by following commands only when preceded by "Simon Says." & Verbal instructions to perform actions, sometimes preceded by "Simon Says." & 10 trials (5 inhibition trials). \\ 
 & & Backward Word Span Task & Working Memory & Measure working memory by repeating verbal sequences in reverse order. & Words presented verbally by the experimenter. & 2 trials per sequence length (2–5 words). \\ \hline

Lertladaluck et al.~\cite{lertladaluck_executive_2024} & 2024 & Stroop Task & Inhibitory Control & Measure cognitive inhibition by responding to shape rather than color. & Shapes (circle, triangle, square) shown in congruent/incongruent colors; response via touchscreen. & 6 blocks (18 trials each: 9 congruent, 9 incongruent). \\ 
 & & Luria Hand Game & Inhibitory Control & Measure inhibition through congruent/incongruent hand gestures. & Experimenter makes gestures (fist/point); child must mimic or do opposite depending on the session. & 8 trials per session (congruent/incongruent). \\ 
 & & Flanker Task & Inhibitory Control & Assess inhibition and selective visual attention by ignoring adjacent distractors. & Fish symbols pointing congruently or incongruently to the central target. & 20 trials (10 congruent, 10 incongruent). \\ \hline

 Lin et al.~\cite{lin_measurement_2019} & 2019 & Shape Stroop Task & Cognitive Flexibility, Inhibitory Control & Measure cognitive flexibility and inhibition by identifying subdominant visual features. & Cards displaying shapes embedded within other shapes. & 24 trials. \\ 
 & & Snack Delay Task & Inhibitory Control & Measure inhibitory control and delay of gratification using edible stimuli. & Transparent cup covering an M\&M candy; delay durations of 10, 20, 30, and 15 seconds. & 4 trials. \\ 
 & & Toy Delay Task & Inhibitory Control & Measure inhibitory control and delay of gratification using non-edible stimuli. & Transparent cup covering "silly bandz"; delay durations of 10, 20, 30, and 15 seconds. & 4 trials. \\ \hline

 Maurer et al.~\cite{maurer_towards_2019} & 2019 & Flanker Task & Inhibitory Control & Measure inhibition by focusing on a target while ignoring distractors. & Visual stimuli: Fish pointing in congruent or incongruent directions. & Block 1: 20 congruent; Block 2: 20 congruent + 20 incongruent. \\ 
 & & Animal Updating Task & Working Memory & Measure working memory updating by recalling the last two animals shown. & Sequences of animals presented for 1900 ms each, followed by a question mark. & 2 test blocks of 5 trials each. \\ 
 & & Advanced Dimensional Change Card Sort (DCCS) & Cognitive Flexibility & Measure cognitive flexibility by sorting cards by color, shape, or a switching rule. & Cards with colors and shapes; instructions change across blocks (e.g., sort by shape if star is present). & Block 1: 10 color; Block 2: 10 shape; Block 3: 20 mixed (shape if star present). \\ \hline

Oeri et al.~\cite{oeri_regulating_2020} & 2020 & Minnesota Executive Function Scale (MEFS) & Cognitive Flexibility, Working Memory, Inhibitory Control & Assess cognitive flexibility, working memory, and inhibition through card sorting tasks. & Cards sorted by changing rules (color vs. shape); increasing difficulty across 7 levels. & 5 trials per part (A and B); adaptive difficulty. \\ 
 & & Flanker Task & Inhibitory Control & Measure response inhibition by identifying the orientation of a central target amidst distractors. & Fish stimuli: congruent (all fish face the same way) or incongruent (target faces opposite). & 24 trials (12 congruent, 12 incongruent). \\ \hline

Papastergiou et al.~\cite{papastergiou_study_2022} & 2022 & Nonverbal Stroop Task & Inhibitory Control & Measure inhibition by identifying arrow directions irrespective of position. & Arrows pointing in different directions; congruent and incongruent blocks. & 3 blocks of 60 trials. \\ & & Backward Digit Span Task & Working Memory & Assess updating by recalling numbers in reverse order. & Auditory presentation of number sequences. & 4–8 numbers per trial, increasing gradually. \\ & & Colour-Shape Task & Inhibitory Control, Cognitive Flexibility & Evaluate shifting between color and shape categorization tasks. & Shapes (circle/triangle) in different colors (red/blue) presented in top or bottom halves. & 3 blocks: 32 trials (single-task blocks) + 64 trials (mixed-task block). \\ \hline

 Schäfer et al.~\cite{schafer_executive_2024} & 2024 & Go/No-Go Task & Inhibitory Control & Assess inhibition by responding to go stimuli and withholding for no-go stimuli. & Animal images (cow = No-Go, chicken/pig/donkey = Go) presented sequentially for 500 ms each. & 60 Go trials, 20 No-Go trials. \\ 
 & & Corsi Blocks Backward Task & Working Memory & Measure working memory by recalling sequences of highlighted blocks in reverse order. & 3x3 grid; fairy images highlight blocks in increasing sequence lengths (1–8 blocks). & 6 sequences per length. \\ 
 & & Flexible Item Selection Task (FIST) & Cognitive Flexibility & Assess cognitive flexibility through categorization shifts based on new shared properties. & Images sharing one or two properties (e.g., color, shape). & 18 one-commonality trials; 9 two-commonalities trials. \\ \hline

Schirmbeck et al.~\cite{schirmbeck_contrasting_2021} & 2021 & Child-Friendly Stroop Task & Inhibitory Control & Assess cognitive inhibition by overriding automatic responses. & Colored shapes, fruits, and vegetables; congruent and incongruent color pairings. & 4 sets (20 items per set). \\ 
 & & Object Span Task & Working Memory & Measure working memory by recalling sequences of objects while performing a secondary task. & Cards with images of objects (e.g., apple, basket); "edible or not?" secondary task. & 5 trials with spans increasing (2–6 objects). \\ \hline

 Traverso et al.~\cite{traverso_relationship_2021} & 2021 & Keep Track Task & Working Memory & Measure WM updating by recalling the last items in designated categories. & Pictures from five categories (animals, sky, fruit, vehicles, clothes). & 6 trials: first 3 with one category; last 3 with two categories. \\ 
 & & Dual Request Selective Task & Working Memory & Assess visual-spatial WM by recalling positions while performing a concurrent task. & Frog pathway on a 4x4 chessboard; clap when frog jumps on a red square. & 10 trials, increasing difficulty (pathway lengths = 2–6). \\ 
 & & Simon Says Task & Inhibitory Control & Measure inhibitory control by performing actions only when prompted by “Simon says.” & Verbal commands with gestures. & 10 trials. \\ 
 & & Preschool Matching Familiar Figure Task & Inhibitory Control & Assess impulsive behavior by comparing target figures to alternatives. & Target figures with five alternatives on a page. & 14 trials. \\ \hline

 Veneziano et al.~\cite{veneziano_individual_2022} & 2022 & Animal Stroop Task & Inhibitory Control & Assess inhibition by naming animal bodies while ignoring faces. & Boards with congruent, control-face, and incongruent animal drawings. & 3 boards (24 items each). \\ 
 & & Classification Task & Cognitive Flexibility & Measure cognitive flexibility by categorizing objects by different criteria. & 9 wooden pieces differing in shape, color, and size; 3 containers for sorting. & 3 sorting trials. \\ 
 & & Local/Global Task & Cognitive Flexibility & Assess flexibility in recognizing global vs. local shapes. & Boards with geometric figures containing smaller shapes within their outlines. & 6 boards. \\ \hline

Xie et al.~\cite{xie_using_2024} & 2024 & Dimensional Change Card Sort (DCCS) & Cognitive Flexibility & Assess cognitive flexibility by sorting cards based on changing rules. & Cards with two attributes (color and shape). & 4 blocks (8 randomized trials per block). \\ 
 & & Go/No-Go Task & Inhibitory Control & Measure inhibition by responding to "Go" stimuli and withholding for "No-Go." & Images of animals (e.g., cows, horses, tigers); dog = "No-Go." & 3 blocks (16 Go, 4 No-Go trials per block). \\ 
 & & Working Memory Task & Working Memory & Measure ability to recall visual-spatial locations. & A monkey appearing on one of six trees; recall the location after the image disappears. & 3 blocks (8 trials each). \\ \hline

Xie et al.~\cite{xie_enhancing_2022} & 2022 & Dimensional Change Card Sort (DCCS) & Cognitive Flexibility & Assess cognitive shifting by sorting cards based on changing rules. & Target cards (red boat, blue rabbit) vs. test cards (blue boat, red rabbit). & 3 sessions, each with pre- and post-switch phases (25s each). \\ 
 & & Go/No-Go Task & Inhibitory Control & Measure inhibitory control by responding to "Go" stimuli and withholding for "No-Go." & Animal images (cow, horse, tiger = "Go"; dog = "No-Go"). & 3 sessions (10 Go trials, 10 No-Go trials per session). \\ 
 & & Missing Scan Task & Working Memory & Measure working memory by identifying missing objects after a brief presentation. & Sets of 4 animals displayed for 10s, disappearing into a "house"; 3 animals reappear. & 3 sessions (5 trials each). \\ \hline

 Zanto et al.~\cite{zanto_digital_2024} & 2024 & Flanker Task & Inhibitory Control & Assess inhibitory control by identifying the central arrow direction amidst distractors. & Five-arrow arrays (congruent/incongruent). & 28 trials (50\% congruent/incongruent). \\ 
 & & Stroop Task & Inhibitory Control & Measure inhibitory control by identifying text color instead of word meaning. & Words representing colors (e.g., "red") displayed in congruent or incongruent colors. & 40 trials (50\% congruent/incongruent). \\ 
 & & Working Memory Task & Working Memory & Measure working memory via reverse Corsi block-tapping task. & Squares illuminated in sequences, requiring participants to recall them in reverse order. & Adaptive trials (set size increases after correct trials). \\ \hline

Zhang et al.~\cite{zhang_recreational_nodate} & 2024 & Stop Signal Task & Inhibitory Control & Evaluate inhibition by stopping responses to a "beep" signal. & Left/right arrows with "beep" stop signal. & 48 trials (16 per condition). \\ 
 & & Letter Memory Task & Working Memory & Assess working memory updating. & Letter sequences displayed, requiring sequential recall. & 8 trials. \\ 
 & & Keep Track Task & Working Memory & Evaluate working memory by tracking recent items in categories. & Visual stimuli from 5 categories (e.g., colors, animals) displayed in octets. & 9 trials. \\ 
 & & Dots Task & Cognitive Flexibility & Assess cognitive flexibility with rule switching. & Sun/moon stimuli requiring same-side or opposite-side responses. & 20 trials. \\ \hline
 
\end{longtable}
\end{landscape}
\end{document}